\documentclass[aps,preprint,amsmath,amssymb,amsfonts,nofootinbib,superscriptaddress]{revtex4-1}
\usepackage{epsfig}
\usepackage{graphicx}
\usepackage{dcolumn}
\usepackage{bm}
\usepackage{amsthm}
\usepackage{amsmath}
\usepackage{color}
\usepackage{natbib} 
\usepackage{graphicx}
\usepackage[export]{adjustbox}
\usepackage{amssymb}
\usepackage{amsmath}
\usepackage{amsfonts}
\usepackage{amssymb}
\usepackage{braket}
\usepackage{xcolor}
\usepackage{subcaption}
\usepackage{float}
\usepackage{tikz}
\usetikzlibrary{arrows.meta, automata,
                positioning,
                quotes}

\usepackage[hidelinks]{hyperref}
\hypersetup{
    colorlinks=false,
    linkcolor=false,
    filecolor=false,      
    urlcolor=false,
    }
\usepackage{float}

\usepackage{cancel}
\usepackage{soul}

\newcommand{\beq}{\begin{equation}}
\newcommand{\eeq}{\end{equation}}
\newcommand{\bea}{\begin{eqnarray}}
\newcommand{\eea}{\end{eqnarray}}

\newcommand{\eq}{\begin{equation}}
\newcommand{\eqx}{\end{equation}}
\newcommand{\eqn}{\begin{eqnarray}}
\newcommand{\eqnx}{\end{eqnarray}}

\newcommand{\f}[2]{\frac{#1}{#2}}

\begin{document}

\title{Turbulence Scaling from Deep Learning Diffusion Generative Models}

\author{Tim Whittaker}%
\affiliation{
Département des sciences de la Terre et de l'atmosphère, Université du Québec à Montréal, Montréal, QC H3C 3P8, Canada
}

\author{Romuald A. Janik}
\affiliation{
Institute of Theoretical Physics and Mark Kac Center for Complex Systems Research,
Jagiellonian University,
ul. {\L}ojasiewicza 11, 30-348 Krak{\'o}w, Poland
}

\author{Yaron Oz}%
\affiliation{
School of Physics and Astronomy, Tel-Aviv University, Tel-Aviv 69978, Israel.
}

\date{\today}

\begin{abstract}
Complex spatial and temporal structures are inherent characteristics of turbulent fluid flows and comprehending them poses a major challenge. This comprehesion necessitates an understanding of the space of turbulent fluid flow configurations.  We employ a diffusion-based generative model to learn the distribution of turbulent vorticity profiles and generate snapshots of turbulent solutions to the incompressible Navier-Stokes equations.
We consider the inverse cascade in two spatial dimensions and generate diverse turbulent solutions that differ from those in the training dataset. We analyze the statistical scaling properties of the new turbulent profiles,
calculate their structure functions, energy power spectrum, velocity probability distribution function and moments of local energy dissipation. All the learnt scaling exponents
are consistent with the expected Kolmogorov scaling. This agreement with established turbulence characteristics provides strong evidence of the model's capability to capture essential features of real-world turbulence.
\end{abstract}

\maketitle
\tableofcontents

\section{Introduction}

Fluid turbulence stands as a profound, unsolved challenge in physics \cite{Frisch}. It manifests as a complex emergent phenomenon, arising from the application of Newton's second law to fluid elements. Extensive research spanning centuries has been dedicated to unraveling the structure of turbulent flows, which encompasses most fluid behaviors in nature across all scales. However, our comprehension of fluid flows in the nonlinear regime remains incomplete.
The study of turbulence holds the promise of shedding light on the principles and dynamics of nonlinear systems characterized by a multitude of strongly interacting degrees of freedom in a far from equilibrium state. 
An intriguing characteristic of turbulence is the phenomenon of scaling, encapsulating the statistical properties and structural complexity of turbulence. Despite significant experimental \cite{Benzi1995OnTS} and numerical \cite{chen_dhruva_kurien_sreenivasan_taylor_2005,Biferale_2019} progress, the precision of available data remains inadequate to definitively distinguish among the various models proposed,  e.g. for the 
anomalous scaling in three-dimensional incompressible fluid turbulence \cite{PhysRevLett.72.336,PhysRevE.63.026307,Eling2015TheAS,Oz:2017ihc}. This emphasizes the need to transition towards an era characterized by ``precision turbulence.''

 Learning capabilities of deep learning algorithms have revolutionized diverse fields, providing a new lens to explore complex systems.
Among various applications, deep learning methods have been increasingly utilized to generate turbulent flows, with several different approaches showing promise. Generative Adversarial Networks (GANs) have been employed to model turbulence in \cite{Drygala_2022,tretiak2022physicsconstrained,Li2023}. The use of Physics-Informed Neural Networks (PINNs), which incorporate physical laws into the learning process, thereby allowing for more accurate predictions in scenarios where the data is sparse or noisy (for a review see \cite{Shu_2023}). In \cite{yang2023denoising} it was proposed to use denoising diffusion models and it was shown to be capable of generating fluid fields from either low resolution or even irregular samples. In a similar vein, super-resolution models, which generate high-resolution output from low-resolution input, have been used for turbulent flow data \cite{fukami_fukagata_taira_2019,ZHOU2022105382}. Super-resolution models effectively bridge the gap between low-resolution measurements and the need for high-resolution reconstructions, making them an ideal tool for the study of turbulence, where fine-scale details can be critical. Another line of research uses diffusion models for generating single particle trajectories in three-dimensional turbulence~\cite{li2023synthetic}.
Deep learning approaches for modeling the temporal evolution in turbulent flows were developed~\cite{mohan2019compressed,king2018deep,
doi:10.1080/14685248.2020.1757685, moghaddam2018deep, li_yang_zhang_he_deng_shen_2020, Buzzicotti2022, PhysRevFluids3104604, lellep_prexl_eckhardt_linkmann_2022}, 
complexity of turbulent versus chaotic snapshots was estimated~\cite{whittaker2022neural},
and other recent use of generative models can be found in \cite{kohl2023turbulent,apte2023diffusion,lienen2023zero}.

Motivated by the promise of deep learning we aim in this work to harness the potential of denoising diffusion probabilistic model (DDPM)s to learn statistical turbulence. The question that we will address is whether deep learning can 
comprehend the properties of turbulence, and whether it can decrease the errors of the training data and make more accurate statistical predictions.
 We will employ a diffusion-based generative model to learn the distribution of turbulent velocity and vorticity profiles and generate snapshots of turbulent solutions to the incompressible Navier-Stokes (NS) equations.
We will consider the inverse cascade in two spatial dimensions, generate diverse new turbulent solutions and analyze the statistical scaling properties of these turbulent profiles.
We will calculate their structure functions, energy power spectrum, velocity probability distribution function and moments of local energy dissipation and show that the learnt scaling exponents
are consistent with the expected Kolmogorov scaling.
The paper is structured as follows: in Sect.~\ref{SEC::Back} we provide a background overview of fluid turbulence scaling, as well as introduce the DDPM. Subsequently, in Sect.~\ref{SEC::Method} we cover our methodology, the specifics of the DDPM, the dataset used, and the approach to model training and evaluation. We then present and discuss the results of our learning experiments in Sect.~\ref{SEC::Results}. We conclude with a discussion of the research findings, their potential applications, and avenues for future research.

\section{Background}
\label{SEC::Back}
\subsection{2D Fluid Turbulence}

The incompressible NS equations provide a mathematical formulation of the fluid flow evolution in $d$ spatial dimensions 
at velocities much smaller than the speed of sound: 
\begin{equation}
\partial_t v^i + v^j\partial_j v^i =
-\partial^i p + \nu \partial_{jj} v^i + f^i,~~~~~~\partial_iv^i = 0  \ ,
\label{NS}
\end{equation}
where $v^i,i=1...d$  is the fluid velocity, $p$ is the fluid pressure, $\nu$ is the kinematic viscosity, and $f^i$ is a random forcing. In the two-dimensional case, it is useful to 
work with the pseudo-scalar vorticity variable
 $\omega = \epsilon_{ij}\partial^iv^j$.

An important dimensionless parameter in the study of fluid flows  is the Reynolds number 
${\cal R}_e = \frac{l v}{\nu}$, where $l$
is a characteristic length scale, $v$ is the velocity difference at that scale, and $\nu$ is the kinematic viscosity. 
The Reynolds number quantifies the relative strength of the
non-linear interaction compared to the viscous term in (\ref{NS}).
When the Reynolds number is of order $10-10^2$ one observes a chaotic 
fluid flow, while when it is $10^3$ or higher,
one observes a fully developed turbulent structure of the flow.
The turbulent velocity field exhibits highly complex spatial and temporal structures and appears to be a random process.
A single realization of a turbulent solution to the NS equations is unpredictable even in the absence of a random force. However, the study of statistical averages reveals a hidden
scaling structure.
Indeed, experimental and numerical  data suggest that turbulent fluid flows exhibit a statistically homogeneous and isotropic steady state at the inertial range of scales $l_v \ll r \ll l_f$, where the distance scales $l_v$ and $l_f$ are determined by the viscosity and driving force, respectively. 

The properties of this statistical structure can be quantified by 
studying statistical averages of fluid observables.
For instance, 
if we denote the velocity of the fluid by $\vec{v}(t,\vec{r})$ then the turbulent behavior can be characterized by the longitudinal structure functions $S_n(r) = \langle (\delta v(r))^n \rangle$ of velocity differences 
$\delta v(r)  = (\vec{v}(\vec{r}) - \vec{v}(0))\cdot \frac{\vec{r}}{|\vec{r}|}$
between points separated by a fixed distance $r$. In the inertial range of scales these correlation functions exhibit a universal scaling law
    $S_n(r) \sim r^{\xi_n}$, 
where the exponents $\xi_n$  are independent of the fluid
details and depend only on the number of spatial dimensions.

In a seminal work \cite{Kolmogorov}, Kolmogorov used the inertial range 
cascade-like behavior (introduced by Richardson)
of incompressible non-relativistic fluids, where large eddies break into smaller eddies in a process where energy is transferred without dissipation. 
Assuming scale invariant statistics for this direct cascade (from large to small length scales), he deduced that $\xi_n = n/3$. 
Thus, for instance, the Fourier transform of $S_2$ gives the energy power spectrum 
that exhibits Kolmogorov scaling:
\begin{equation}
    E(k) \sim k^{-\frac{5}{3}} \ .
    \label{EK}
\end{equation}
It is established numerically and experimentally that Kolmogorov linear scaling is corrected by intermittency in the direct cascades. Kolomogorov scaling seems to hold in the two-dimensional inverse cascade, where the energy flows
from the UV to the IR and the inertial range holds for $r \gg l_f$.

A correspondence can be established between two-dimensional scaling and the local energy dissipation, 
$\epsilon(x) = \frac{\nu}{2} \left(\partial_i v^j + \partial_j v^i \right)^2$.
Taking the normalized local spatial average of the energy dissipation over a ball with a $d$-dimensional radius, $r$, denoted as $B_d(r)$, and centered around a point $x$:
\begin{equation}
\epsilon_r(x) = \frac{1}{Vol(B_d(r))} \int_{|x'-x| \leq r} d^d x' \epsilon(x') \ , \label{mea}
\end{equation}
the ensemble averages according to the K41 theory satisfy:
\begin{equation}
\langle \epsilon_r^n \rangle \sim r^{\tau_n},~~~ \tau_{\frac{n}{3}} = \left( \xi_n - \f{n}{3} \right)  \  . \label{eq:led}
\end{equation}
In the two-dimensional inverse cascade case studied in the present paper we expect $\tau_n=0$.

\subsection{Diffusion Generative Models}

A DDPM~\cite{ho2020denoising} is a powerful probabilistic generative framework that has shown success in transforming and generating images (see \cite{yang2023diffusion} for a review), by progressively injecting noise into the original data and subsequently reversing the process during sample generation.
The process begins with the original data and perturbs it leading to noisy data. The goal is to transform the data distribution into a simple prior distribution. Given a data distribution ${\bf x_0} \sim q({\bf x_0})$, a sequence of random variables (RV), ${\bf x_0},{\bf x_1},...{\bf x_T}$, are generated from a Markov process with transition kernel $q({\bf x_t}|{\bf x_{t-1}})$. In a DDPM, the kernel is designed to transform the distribution $q({\bf x_0})$ into a Normal distribution,
\begin{equation}
    q({\bf x_t}|{\bf x_{t-1}}) = \mathcal{N}({\bf x_t};\sqrt{1-\beta_t}{\bf x_{t-1}}, \beta_t{\bf 1}) \ ,
\end{equation}
where $\beta_t \in (0,1)$ is a hyper-parameter. The joint distribution of the RVs can be factorized using the Markov property and the chain rule to get,
\begin{equation}
    q({\bf x_1},...{\bf x_T} | {\bf x_0}) = \prod_{t=1}^T q ({\bf x_t}|{\bf x_{t-1}}) \ .
\end{equation}
Now using the Gaussian kernel we can marginalize the joint distribution, we get 
\begin{equation}
    q({\bf x_t} | {\bf x_0}) = \mathcal{N}({\bf x_t};\mu_t{\bf x_{0}},\sigma_t{\bf 1}) \ ,
\end{equation}
where $\mu_t = \sqrt[]{1-\beta_t}$, and $\sigma_t = 1-\prod_{s=0}^t \mu_s^2$. In this sense the forward process transforms the data distribution into a Normal distribution.

When generating new data samples using DDPMs, an unstructured noise vector is generated from the prior distribution. Since the prior distribution is typically chosen as a simple Gaussian distribution, obtaining this noise vector is straightforward. To gradually remove the noise from this noise vector and generate meaningful data, a learnable Markov chain operates in the reverse time direction. The reverse Markov chain consists of transition kernels parameterized by deep neural networks (a U-net in our case). These transition kernels are designed to undo the perturbations caused by the forward process and recover the original data
\begin{equation}
    p_{\theta}({\bf x_{t-1}}|{\bf x_t}) = \mathcal{N}({\bf x_{t-1}}; \mu_\theta({\bf x_t},t), \sigma_\theta({\bf x_t},t)) \ ,
\end{equation}
where $\theta$ are the model parameters tuned during the training process. The training process consists of minimizing the distance between the reverse process joint distribution $p_\theta({\bf x_0},{\bf x_1},...,{\bf x_T})$ and the forward process $q({\bf x_0},{\bf x_1},...,{\bf x_T})$. To this end, the usual variational bound on negative log likelihood is optimized:
\begin{align*}
\mathbb{E} [-\log p_{\theta} (x_0)] \leq \mathbb{E}_{q} \left[ -\log \frac{p_{\theta} (x_{0:T})}{q(x_{1:T} | x_0)} \right] \\
= \mathbb{E}_{q} \left[ -\log p(x_T) - \sum_{t \geq 1} \log \frac{p_{\theta} (x_{t-1} | x_t)}{q(x_t | x_{t-1})} \right] \\
= \mathbb{E}_{q} \left[ -\log p_{\theta} ({\bf x}_0) + \sum_{t=1}^T \text{KL}(q({\bf x}_t | {\bf x}_{t-1}) \| p_{\theta}({\bf x}_{t-1}|{\bf x}_t)) \right]\\
=:\mathcal{L}    
\end{align*}
For normal Gaussian distributions, the KL divergence: 
\begin{equation}
\text{KL}(q({\bf x}_t | {\bf x}_{t-1}) \| p_{\theta}({\bf x}_{t-1}|{\bf x}_t)) = \frac{1}{2} \left( \text{tr}(\sigma_{\theta}^{-2} \sigma_{q,t}^2{\bf I}) + (\mu_{\theta} - \mu_{q,t})^\top \sigma_{\theta}^{-2} (\mu_{\theta} - \mu_{q,t}) - d + \log \frac{|\sigma_{\theta}^2{\bf I}|}{|\sigma_{q,t}^2{\bf I}|} \right),
\end{equation}
where \( d \) is the dimensionality of the Gaussian distributions, and for each time step \( t \).  The loss function $\mathcal{L}(\theta)$ is thus composed of the expected negative log likelihood of the data under the model and the sum of KL divergences across all timesteps, which measures the discrepancy between the forward and reverse transition probabilities.

\section{Methodology}
\label{SEC::Method}
\subsection{Model and Architecture}

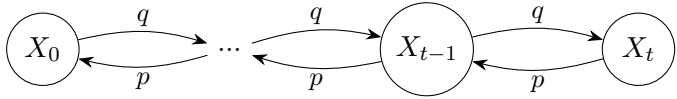
\begin{figure}[ht]
\begin{tikzpicture}[
    node distance = 29mm and 17mm,
every edge/.style = {draw, -{Stealth[scale=1.2]}, bend left=15},
every edge quotes/.append style = {auto, inner sep=2pt, font=\footnotesize}]

\node (n1)  [state] {$X_0$};
\node[draw=none] (n12)  [right=of n1]   {$...$};
\node (n2)  [state,right=of n12]   {$X_{t-1}$};
\node (n3)  [state, right=of n2]   {$X_{t}$};

\path   (n1)    edge ["$q$"] (n12)
        (n12)    edge ["$p$"] (n1)
                edge ["$q$"] (n2)
        (n2)    edge ["$p$"] (n12)
                edge ["$q$"] (n3)
        (n3)    edge ["$p$"] (n2);
\end{tikzpicture}
    \caption{Forward noising and backward denoising Markov chains.}
    \label{fig:UNET}
\end{figure}

We adopt a fairly standard diffusion model architecture based on the U-Net with components from the \texttt{hugginface diffusers} library \cite{von-platen-etal-2022-diffusers}. Refer to Figure \ref{fig:UNET} for a diagram of the Markov chain model, where the U-Net architecture is employed to parameterize the p kernel. The input/output image size is $256\times 256$ with 7 downsampling and upsampling blocks. The 6th downsampling and the corresponding 2nd upsampling block have in addition spatial self-attention. The respective number of channels is $128, 128, 256, 256, 512, 512, 1024$, which is  comparable to diffusion models generating real-world images. The overall number of trainable parameters is $28.22\times 10^7$.

\subsection{Training Datasets}
\label{s.numerics}
We generated turbulent data by solving NS equations on a uniform spatial grid spanning a domain of \(L_x = L_y = 2\pi\)
as in \cite{whittaker2022neural}. Initialized with \(v = (0,0)\), this system underwent numerical evolution with periodic boundary conditions. To drive turbulence, we applied a divergence-free, statistically homogeneous, and isotropic Gaussian random forcing function within an annulus in Fourier space centered at \(k_{f}\). For numerical reasons, the second-order viscous term in eq.~(\ref{NS}) was replaced with a hyperviscous term, as discussed in \cite{2dTurbReview}. We use a dealiased spectral method code with Crank-Nicolson time stepping \cite{specmeth}.

\begin{figure}[ht]
    \centering
    \includegraphics[width=0.45\textwidth]{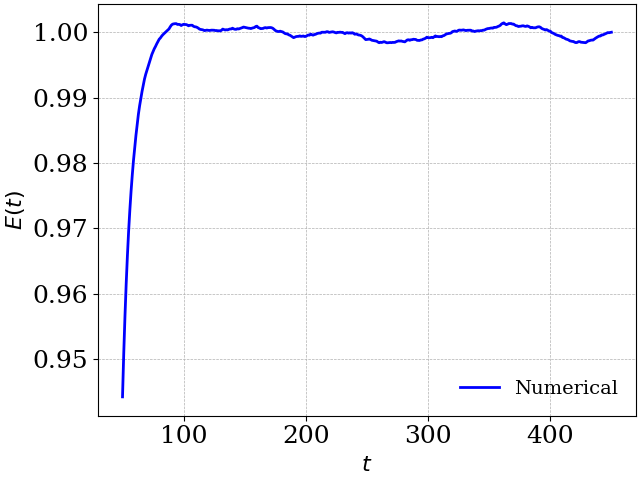}
    \includegraphics[width=0.45\textwidth]{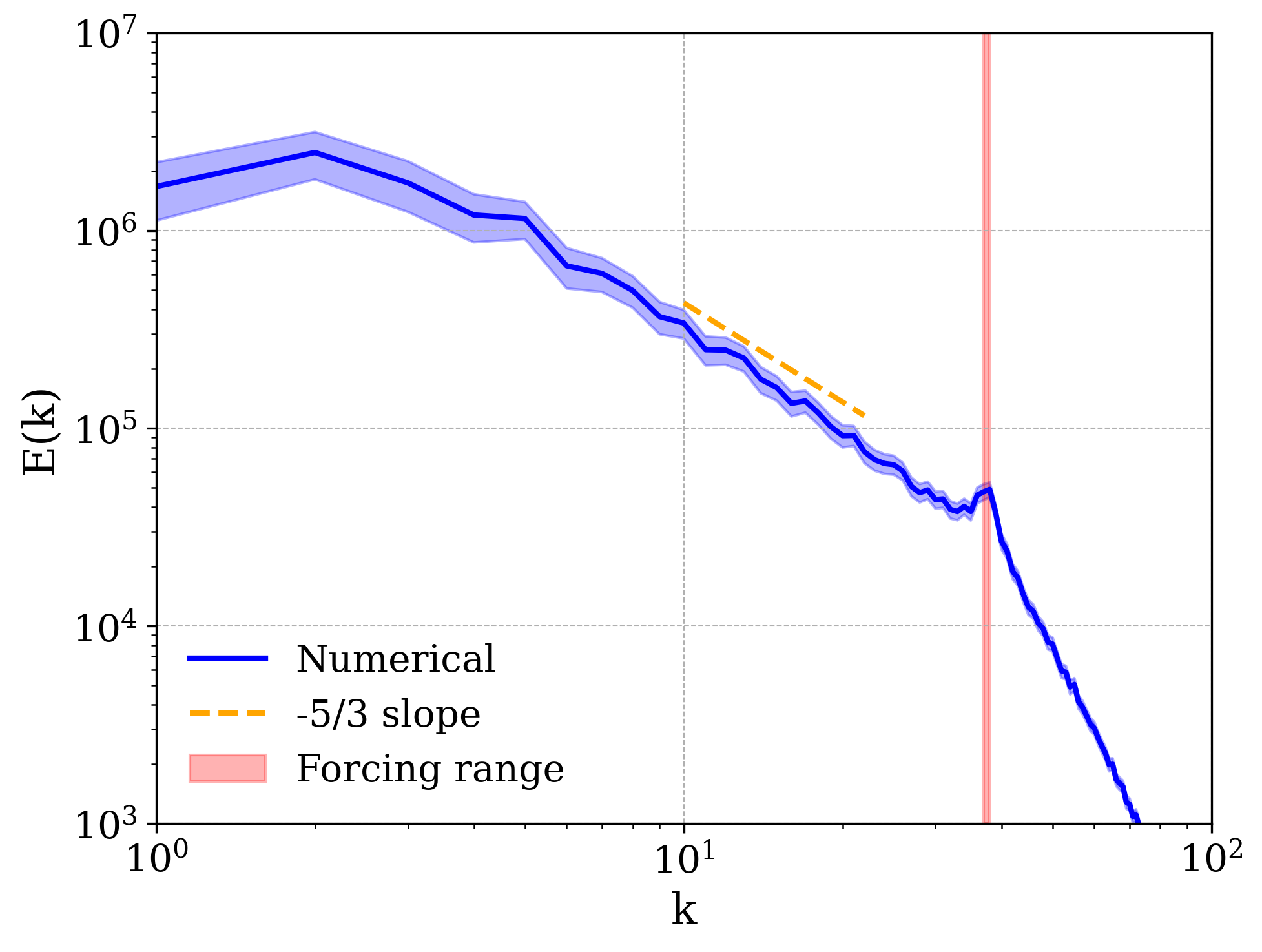}
    \caption{Left: Evolution in time of the fluid energy, highlighting the attainment of steady state. Right: Energy power spectrum showcasing the 
    \(-5/3\) scaling with standard deviation in the shaded regions, indicative of turbulence state in the inertial range.}
    \label{fig:sim_properties}
\end{figure}

Our dataset consists of \(5000\)  snapshots from an ensemble of ten simulations, each with a resolution of \(512 \times 512\) pixels and a forcing parameter of \(k_f \sim 40\). Upon evolution, the system reached a steady state, as shown in the left panel of Fig.~\ref{fig:sim_properties}. At this state, we observed the  \(-5/3\) scaling of the energy power spectrum, marking the system's transition to turbulence. This observation is illustrated in the right panel of Fig.~\ref{fig:sim_properties}. This quantity is computed as the mean over the ensemble and time slices. We note that 2D turbulence has the possibility of producing double cascades as shown in \cite{doi:10.1063/1.1762301} and numerically produced in \cite{PhysRevLett.81.2244,PhysRevE.82.016307}, though we do not generate the direct cascade in our simulations. 

The inertial range of the simulations was determined by initially calculating the third-order structure function, which is expected to exhibit a scaling behavior such that $S_3(r)\sim r$. We designate the inertial range as the interval where this scaling relation fits optimally.

Due to the DDPM's memory constraints, which restricts it to displaying and generating images of \(256 \times 256\) resolution, we downscaled our simulation data. Additionally, we converted the data values from floating-point numbers to integers in the 0-255 range. This resizing was performed using bilinear interpolation onto a \(256\times 256\) grid.

\subsection{Training Procedure}

The diffusion model is trained for 50 epochs with a batch size of 16, an AdamW optimizer \cite{loshchilov2019decoupled}, base learning rate $1e-4$ and cosine learning rate scheduler with a warm-up of 500 steps. Gradient norm is clipped to 1.0. Automatic mixed precision is employed.

The training data comprises $5000$ $256 \times 256$ vorticity images. During training, 
each image is
rotated by mutliples of 90 degrees and/or mirror reflected. For the investigation of memorization presented in section~\ref{s.memorization}, this augmentation procedure is turned off and the 
image is
used without any rotation or reflection.

\section{Results}
\label{SEC::Results}
\subsection{Generated Samples}

\begin{figure}[ht]
    \centering
    \includegraphics[width=0.45\textwidth]{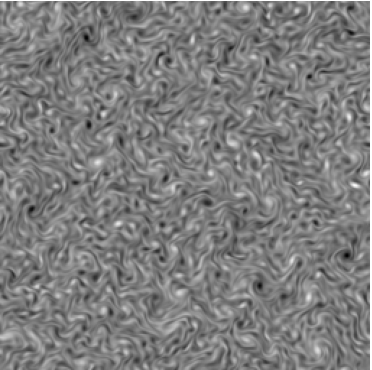}
    \includegraphics[width=0.45\textwidth]{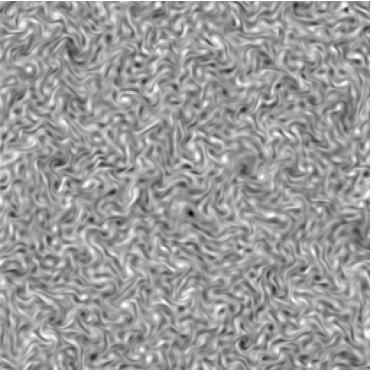}
    \caption{Sample images from the training set (left), and generated by the diffusion model (right).}
    \label{fig.samples}
\end{figure}

In Fig.~\ref{fig.samples}, we show a sample $256\times 256$ image from the vorticity profiles generated using the numerical simulations described in section~\ref{s.numerics} and an image generated by the trained diffusion model.
The generated images look very similar to the real ones, so as to be basically indistinguishable by eye. We observe only a relatively large variation in overall lightness of the generated images. This might be due to accumulating overall systematic shifts in the generation procedure (which requires iterative evaluation of the neural network). 
However, as the precise linear mapping between pixel intensities and values of vorticity is not essential for extracting the statistics of turbulence (and also varies in our training set), we did not attempt to ameliorate this behaviour.
Indeed, as we show in subsequent analysis, various normalization independent quantitative characteristics of turbulent vorticity profiles are very well reproduced in the images generated by the diffusion model.

\subsection{A Test for Memorization}
\label{s.memorization}

\begin{figure}[ht]
    \centering
    \includegraphics[width=0.44\textwidth]{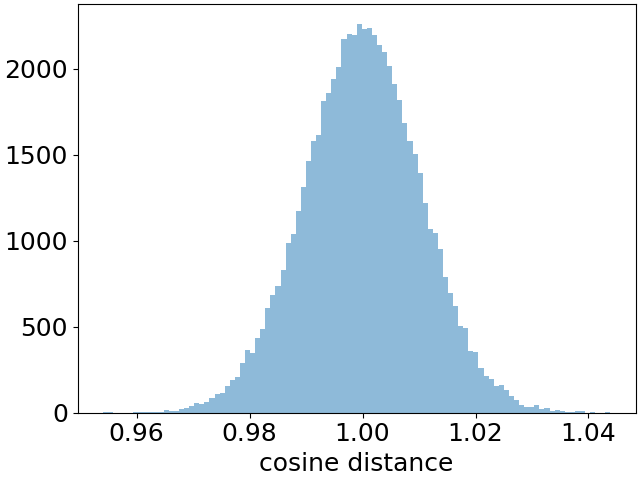}\hfill
    \raisebox{0.9cm}{\includegraphics[width=0.84\textwidth]{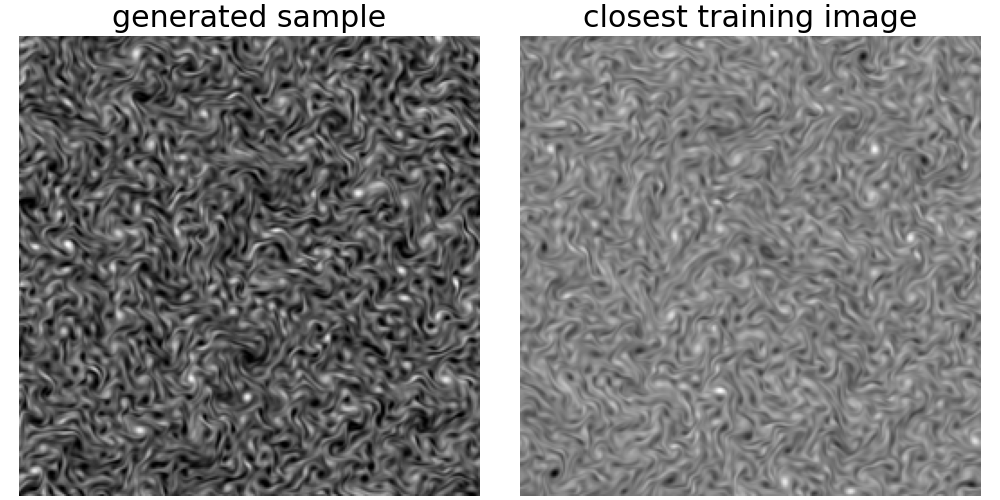}}
    \caption{The histogram of cosine distances between 16 generated samples and the 5000 training images (left) and a pair of the most similar sample and training image (right).}
    \label{fig.similarity}
\end{figure}

An important requirement for the application of neural network models for generating new samples of turbulence profiles is that the generated samples are genuinely new and not just memorized images from the training data. Judging by experience with generative neural network and real world images we do not expect this to be a problem. Nevertheless, we perform a quantitative test, as it is much more difficult to judge by eye the similarity of turbulence profiles in contrast to e.g. celebrity faces.

In order to measure the similarity of generated images to the training data we use the standard cosine distance between the vectors of pixel intensities obtained by flattening the 2D images:
\eq
cosine\ distance(v_1, v_2) = 1 - \f{v_1\cdot v_2}{|v_1| |v_2|}
\eqx
Such a pixelwise comparison is appropriate as a test of memorization.
To perform this experiment we turned off image augmentation and trained the diffusion model 
without any subsequent reflection or rotation. This 
somewhat
reduced the diversity of the training dataset, making the danger of overfitting (or memorization) more acute.

We generated a batch of 16 images and evaluated their cosine distance to all the 5000 training images. The histogram is shown in Fig.~\ref{fig.similarity} (left) and we see that all the distances are centered around 1 -- which is the extreme distance in this metric. We also identified the most similar pair of generated and training images which we show in Fig.~\ref{fig.similarity} (right). The two images are globally clearly different.
Hence, the diffusion model indeed generates genuinely novel samples.

\subsection{Inverse Cascade}
\begin{figure}[ht]
    \centering
    
    \begin{subfigure}{0.45\textwidth}
        \centering
        \includegraphics[width=\textwidth]{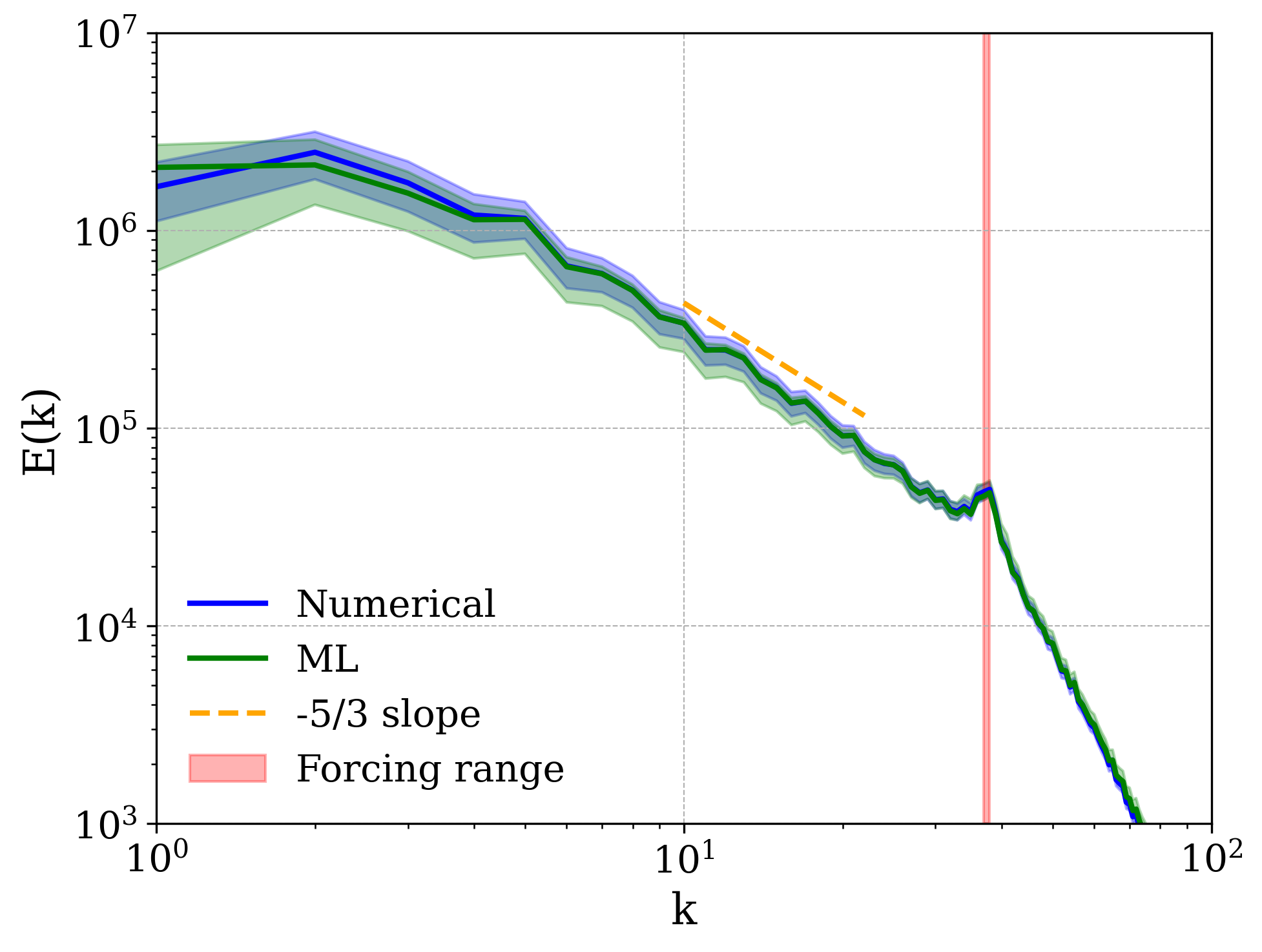}
        \caption{}
        \label{fig:Ek_result}
    \end{subfigure}
    \hfill
    \begin{subfigure}{0.45\textwidth}
        \centering
        \includegraphics[width=\textwidth]{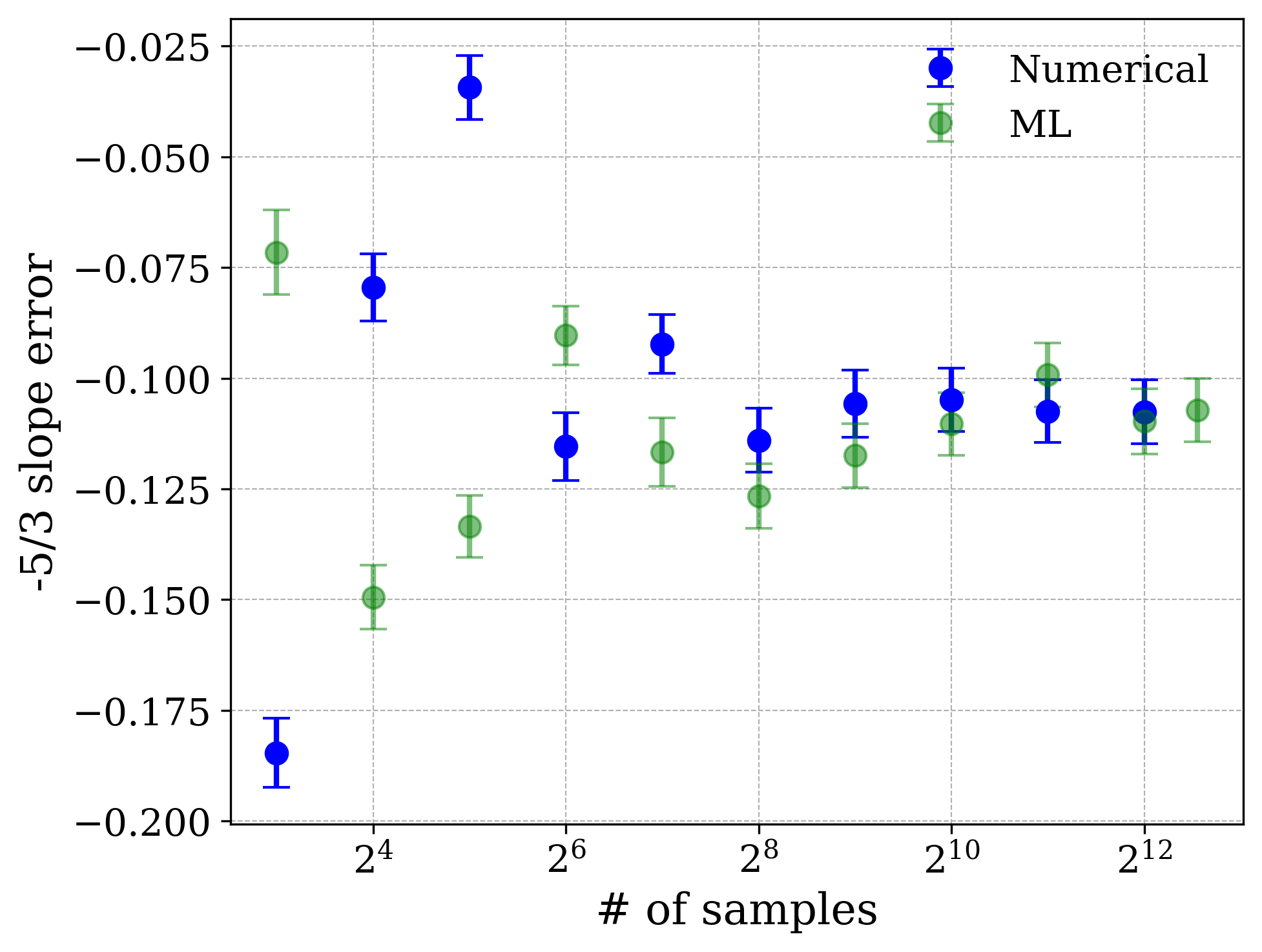}
        \caption{}
        \label{fig:slope_error}
    \end{subfigure}
    
    \begin{subfigure}{0.45\textwidth}        \centering
        \includegraphics[width=\textwidth]{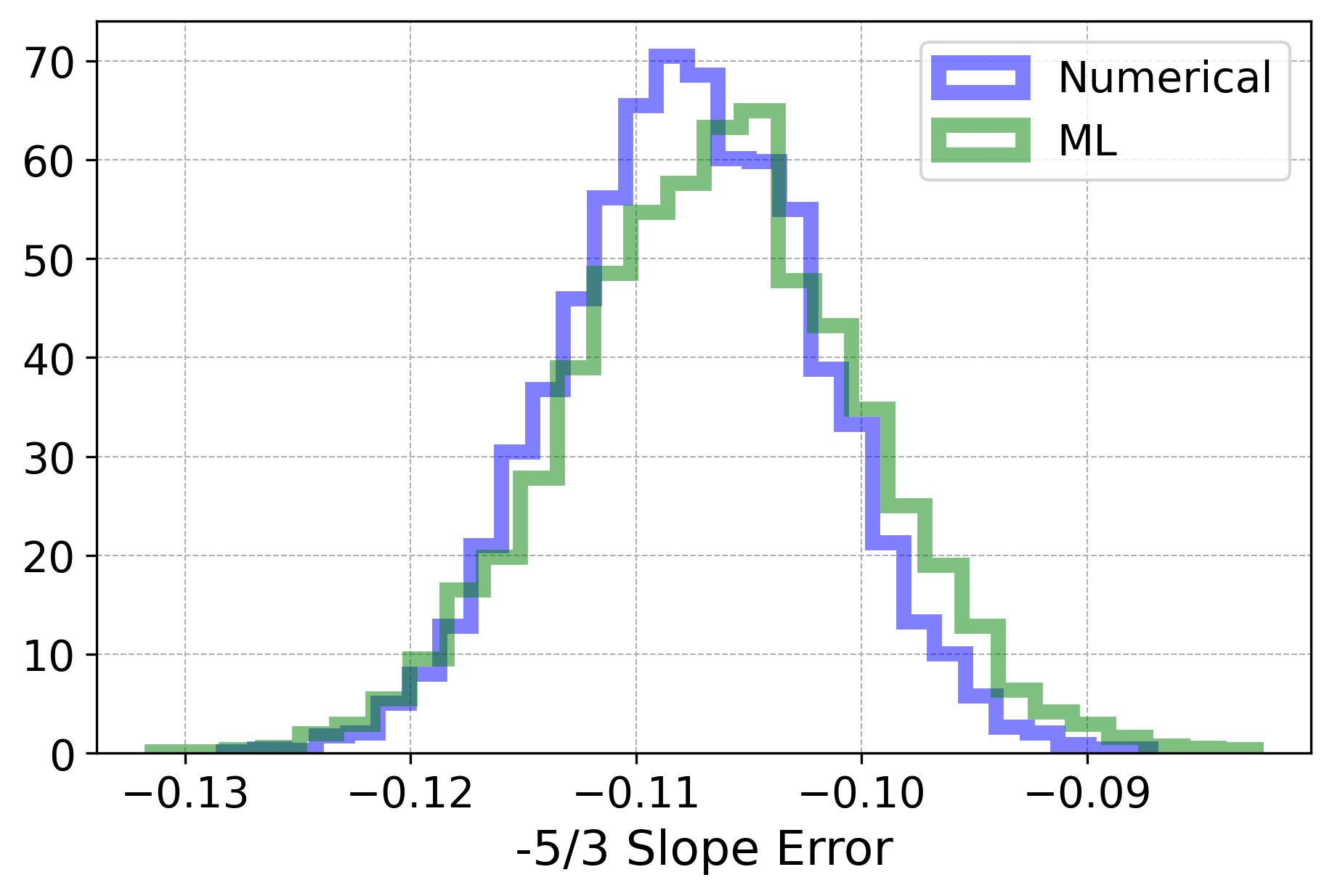}
        \caption{}
        \label{fig:slope_error_bootstrap}
    \end{subfigure}

    \caption{\textbf{(a)} Energy cascade from the numerical simulation and the DDPM, emphasizing the $-5/3$ slope in the inertial range with a noted discrepancy at lower wavenumbers. The standard deviation is shown in the shaded regions. The vertical line indicates the forcing scale.  \textbf{(b)} Measured slope error with standard errors of the fits across varying set sizes, revealing the pronounced influence of the selected range. \textbf{(c)} Measured slope errors using bootstraping.}
    \label{fig:combined_insights}
\end{figure}

We begin by assessing the capability of the DDPM in replicating characteristics of the energy cascade, using numerical simulations as a benchmark. The dataset features an inverse cascade, characterized by a $-5/3$ scaling within the inertial range. In Figure~\ref{fig:Ek_result}, both the mean and variance of the energy spectrum of the ensemble derived from the DDPM closely match those of the numerical simulation, particularly evident in the $-5/3$ scaling within the inertial range.
Notwithstanding this agreement, discrepancies are observed at the lower wavenumbers.

For a more comprehensive analysis, slopes within the inertial range were examined across varying sample set sizes to quantify deviation from the $-5/3$ scaling. The scaling is measured within the range highlighted in Fig.~\ref{fig:Ek_result} by the orange slope. The measured error for the numerical simulation and the DDPM is depicted in Fig.~\ref{fig:slope_error} with the standard error of the linear fits. We employed bootstrapping to gain a more granular understanding of the errors. This involved 5000 iterations, each with sample sizes of 1000. The findings from this exercise are depicted in Fig.~\ref{fig:slope_error_bootstrap}. We find good agreement between the DDPM and numerical distribution indicating the machine has successfully learnt the distribution.

\subsection{Structure Function}
To delve deeper into the statistics of the turbulent flows, we turn our attention to velocity structure functions, defined as:
\begin{equation}
\label{e.Cn}
 S_n(r) = \langle (\delta v(r))^n \rangle \ ,
\end{equation}
where averaging occurs over all positions $x$ within a specific velocity profile image and then extends to various turbulence realizations. 
The energy power spectrum is the Fourier transform of $S_2$, and as discussed in the background section, we expect $S_n(r) \sim r^{n/3}$.

Our primary data sources, namely the images from the DDPM and the numerical simulations, provides vorticity. To correlate this with our structure functions, we first derive the velocities from these vorticity profiles, as elaborated in App.~\ref{App:vort-vel}. When deriving this observable from images, whether they are machine-generated or sourced from the diffusion model, it's crucial to acknowledge a linear transformation between pixel intensities and vorticity:
\begin{equation}
intensity(x) = \alpha \cdot v(x) + \beta \ .
\end{equation}
This transformation might differ across images. As such, to ensure consistency, we normalize the above correlator by $S_n(r=1px)$ where $r=1$ pixel.

\begin{figure}[ht]
    \centering
    \includegraphics[width=0.41\textwidth]{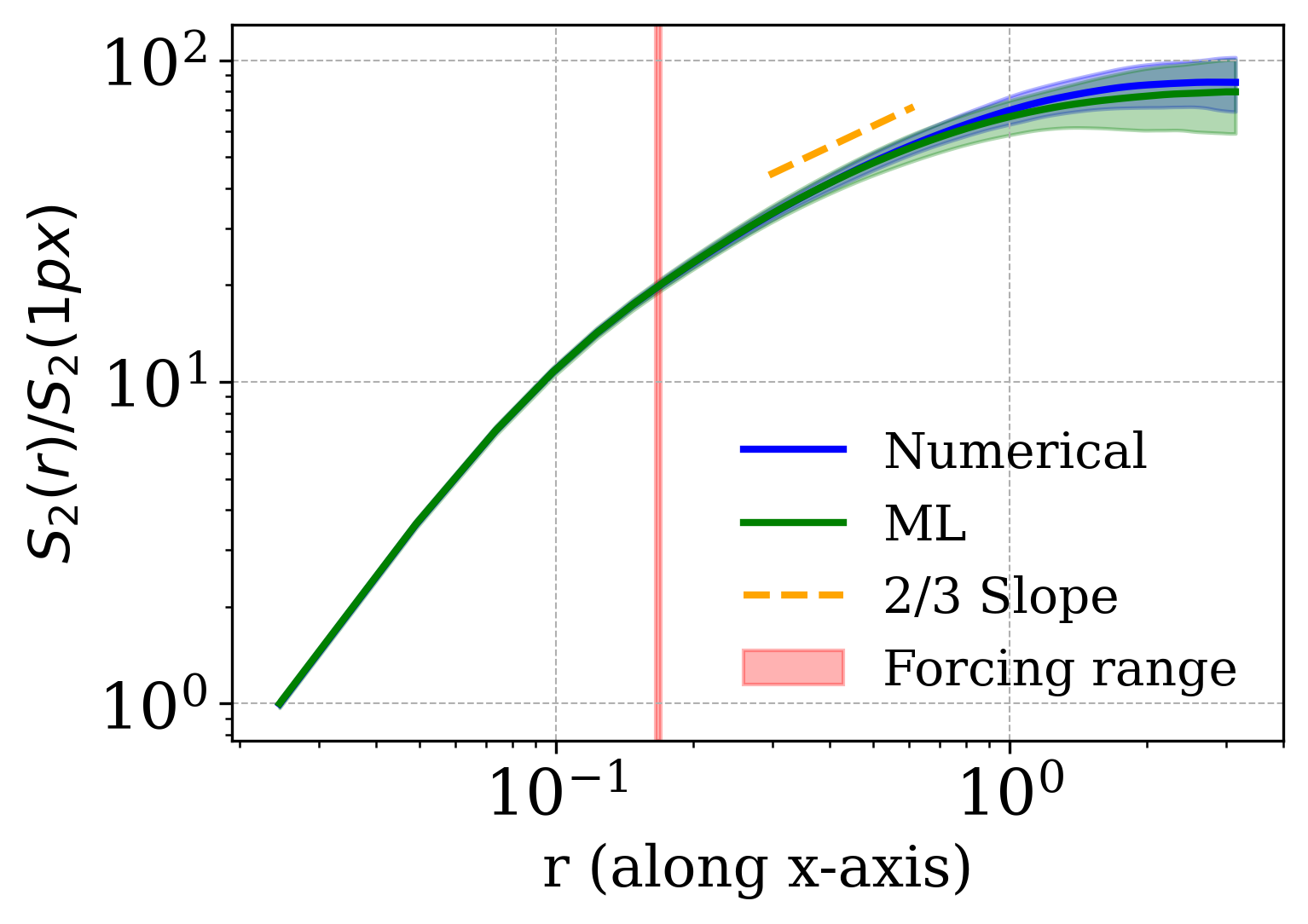}
    \includegraphics[width=0.45\textwidth]{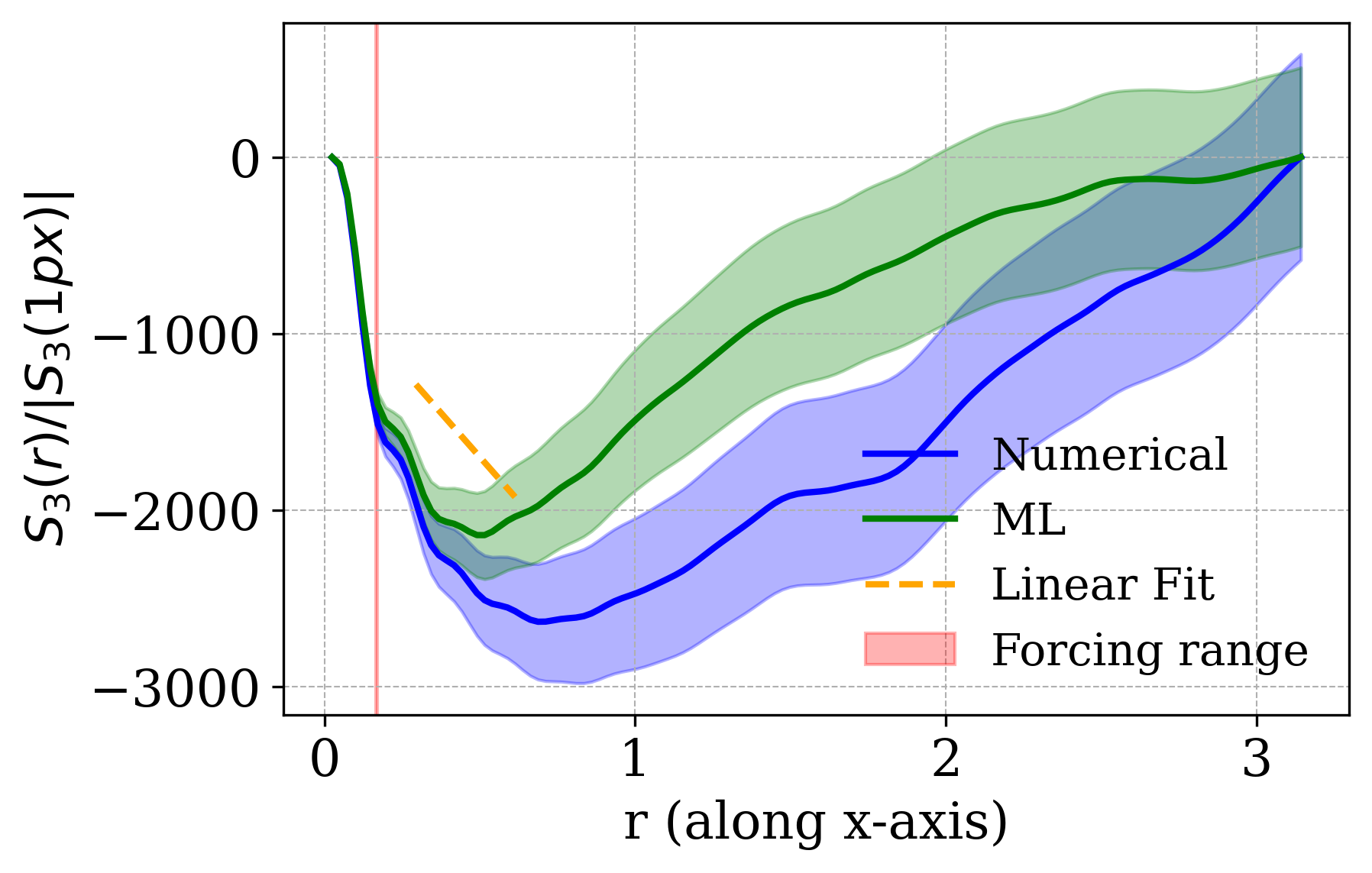}
    \includegraphics[width=0.45\textwidth]{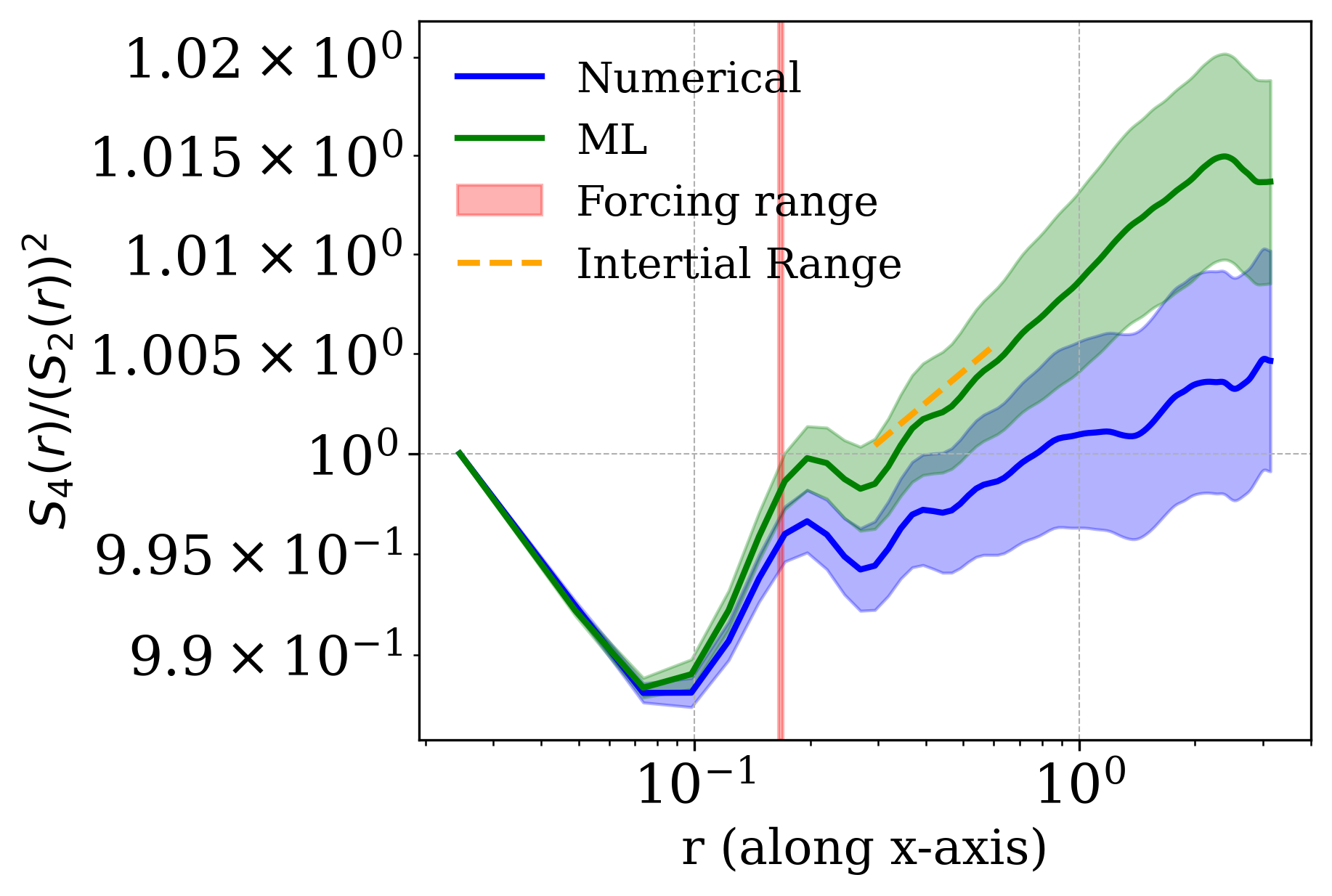}
    \caption{Second (top left) and third (top right) order structure function for both simulations and the machines results with standard deviation in the shaded regions. The dashed orange line shows the 2/3 slope and linear slope for the second and third moments respectively. Bottom left: $S_4(r) / (S_2(r))^2$ is consistent with no intermittency at the inertial range of scales, as evidenced by a nearly flat slope, $ \sim 0.0003$, in this range.}
    \label{fig:structure}
\end{figure}

\begin{figure}[ht]
    \centering
    \includegraphics[width=0.45\textwidth]{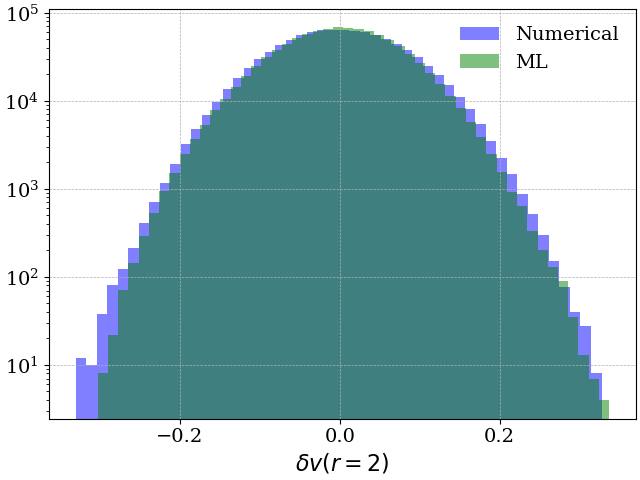}
    \includegraphics[width=0.45\textwidth]{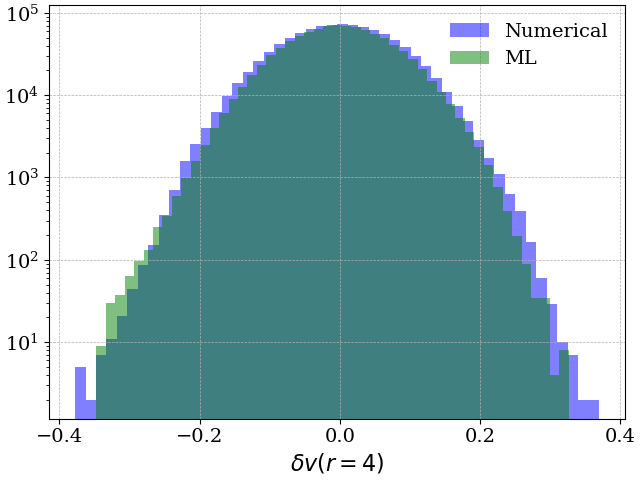}
    \caption{The distribution of $\delta v$ at $r=2$ (left) and $r=4$ (right). We compute the KL divergence between the distributions and find for $r=2$, $D_{KL} = 0.033$ while for $r=4$, $D_{KL}=0.007$}
    \label{fig:dv_dist}
\end{figure}

In Fig.~\ref{fig:structure}, we present a comparative analysis of:
\begin{equation}
\frac{1}{n} \log \left\langle \frac{S_n(r)}{S_n(1{px})} \right\rangle,
\end{equation}
for $n=2,3$. 
We see that both the numerical and DDPM results agree fairly well. We evaluate the intermittency parameter, defined as $S_4(r)/(S_2(r))^2$, and observe a minor slope, suggesting the anticipated independence from $r$. The standard deviation for $n=2,3,4$ is also computed, demonstrating agreement between the numerical and DDPM ensemble results. Additionally, we compare the probability distribution functions of $\delta v$ (refer to Fig.~\ref{fig:dv_dist}) across two distances and find that the resulting distributions are nearly indistinguishable.

\subsection{Local Energy Dissipation}
\begin{figure}[ht]
    \centering
    \includegraphics[width=0.45\textwidth]{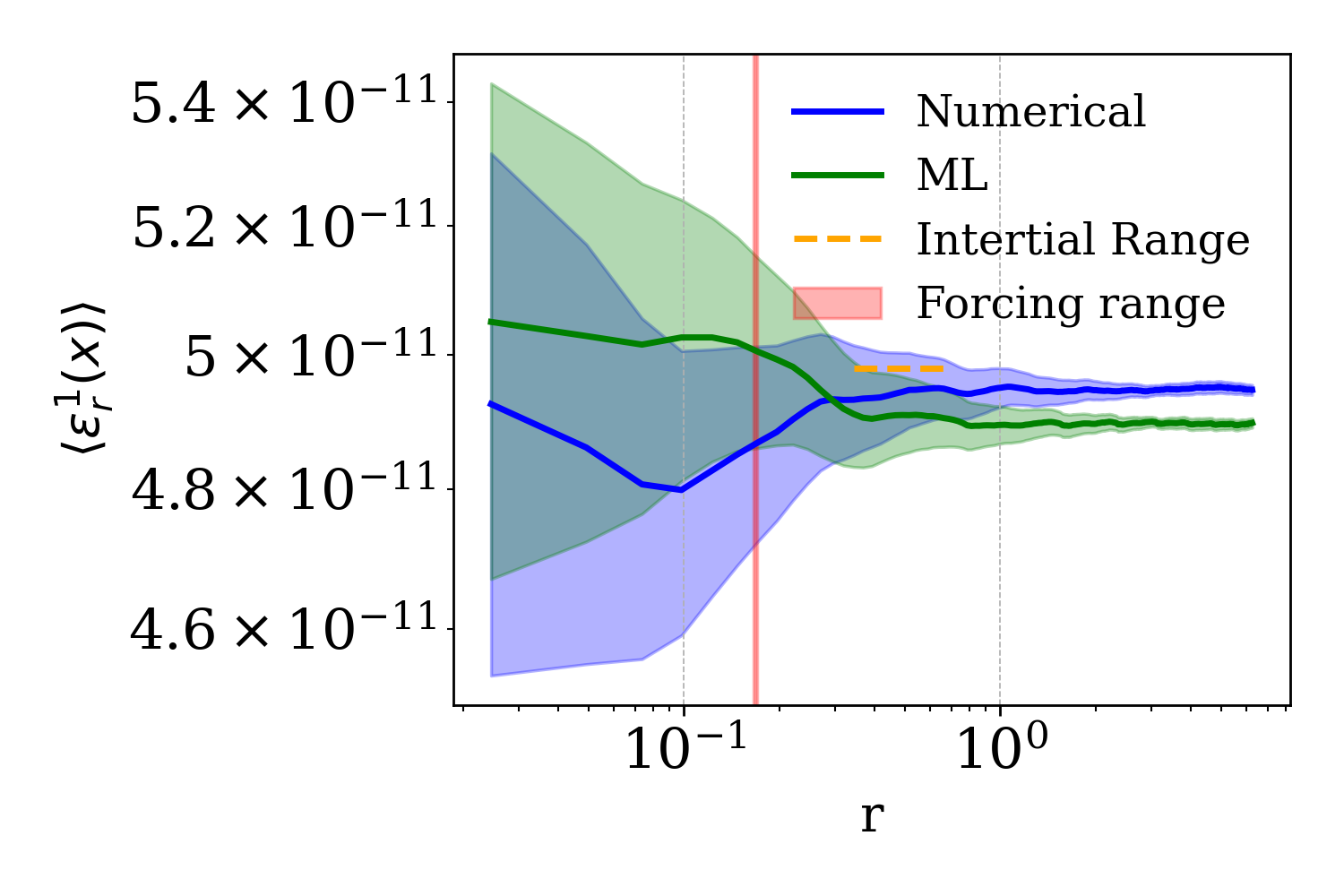}
    \caption{Energy dissipation. Eq.~\ref{mea} estimated with Monte Carlo over 150 snapshots with the standard deviation of the ensemble is shown in the shaded region. We observe that both datasets produce a similar result in the inertial range. The local energy dissipation is independent of $r$ as expected from eq.~\ref{eq:led}.}
    \label{fig:energy_diss}
\end{figure}

Finally, we assessed the local energy dissipation as detailed in eq.~\ref{mea} for $n=1$. This assessment was conducted over a sample of 150 images with random $x$ sampling. In Fig.~\ref{fig:energy_diss}, a comparison of the numerical simulation and the DDPM outcomes is presented. Notably, there's a consistent agreement between both datasets within the inertial range, aligning with theory (eq.~\ref{eq:led}). We observe a significant variance stemming from the limited number of samples available. However, the variance gradually diminishes as we approach the inertial range.

\section{Discussion}

In this paper we have employed a generative diffusion neural network model (DDPM) to generate snapshots of 2D turbulence. The DDPM was trained on vorticity profiles extracted from direct numerical simulations of the incompressible NS equations in two spatial dimensions. We verified that the generated samples exhibit the key statistical properties of turbulence, namely the -5/3 scaling of the energy cascade (Fig.~\ref{fig:combined_insights}), the behaviour of the second, third and fourth structure function (Fig.~\ref{fig:structure}) as well as supporting the conjectured behaviour of energy dissipation (Fig.~\ref{fig:energy_diss}). In all cases, the statistics of the generated data follow rather closely the properties extracted from direct  numerical simulations. 

These results indicate that deep learning diffusion generative models may serve as a useful tool for learning statistics of turbulence and creating proxy independent flow profiles, which can be used to increase statistics for analyzing turbulence. We note that the diffusion models essentially work ``out of the box'' and can generate very realistic turbulent profiles. This is in contrast to other standard deep learning generative approaches such as Generative Adversarial Networks (GAN) and Variational Auto-Encoders (VAE). Prior to this work we tested a state of the art GAN network (StyleGAN) and some variants of variational autoencoders but could not attain sufficiently realistic vorticity profiles. The diffusion models seem to work much better in this respect. We have to emphasize, however, that whether a particular generative model works better or worse is really an empirical question given our current knowledge of deep learning, and may depend on the details of the specific use case. Indeed there are examples where GANs seem to work quite well for convective turbulence~\cite{heyder2023generative}.

We note that the flexibility of the diffusion models in mimicking the training data may be sometimes a two-edged sword. The statistics of the generated data mimic, by construction, the statistics of the training data including all \emph{non-universal} particularities of the data, like behaviour away from the inertial range etc. or any deviations like non-fully developed turbulence.
Therefore, the generative model may not be able to cure possible systematic deficiencies of the data used for training. This can
set a high bar for the quality of the simulation data used for training the generative diffusion model w.r.t to the physical properties that we are interested in studying. 

Our application of the diffusion model to the study of statistical turbulence is limited by the 
dataset of solutions to the Navier-Stokes equations. In particular, it would be desirable to include turbulent fluid solutions 
at higher Reynolds numbers which will enlarge the inertial range of scales.

\noindent{}{\bf Acknowledgments.} 
This research was enabled in part by support provided by Calcul Québec (calculquebec.ca) and the Digital Research Alliance of Canada (alliancecan.ca). R.J. was supported by the research project \textit{Bio-inspired artificial neural networks} (grant no. POIR.04.04.00-00-14DE/18-00) within the Team-Net program of the Foundation for Polish Science co-financed by the European Union under the European Regional Development Fund and by a grant from the Priority Research Area DigiWorld under the Strategic Programme Excellence Initiative at Jagiellonian University. The work of Y.O is supported by the ISF center of excellence and the 
U.S-Israel Binational Science Foundation.



\bibliographystyle{JHEP}
\bibliography{sample} 

\providecommand{\href}[2]{#2}\begingroup\raggedright\begin{thebibliography}{10}

\bibitem{Frisch}
U.~Frisch, \emph{Turbulence: The Legacy of A. N. Kolmogorov}, Cambridge
  University Press (1995),
  \href{https://doi.org/10.1017/CBO9781139170666}{10.1017/CBO9781139170666}.

\bibitem{Benzi1995OnTS}
R.~Benzi, S.~Ciliberto, C.~Baudet and G.R.~Chavarria, \emph{On the scaling of
  three-dimensional homogeneous and isotropic turbulence}, {\emph{Physica D:
  Nonlinear Phenomena} {\bfseries 80} (1995) 385}.

\bibitem{chen_dhruva_kurien_sreenivasan_taylor_2005}
S.Y.~Chen, B.~Dhruva, S.~Kurien, K.R.~Sreenivasan and M.A.~Taylor,
  \emph{Anomalous scaling of low-order structure functions of turbulent
  velocity}, \href{https://doi.org/10.1017/S002211200500443X}{\emph{Journal of
  Fluid Mechanics} {\bfseries 533} (2005) 183–192}.

\bibitem{Biferale_2019}
L.~Biferale, F.~Bonaccorso, M.~Buzzicotti and K.P.~Iyer, \emph{Self-similar
  subgrid-scale models for inertial range turbulence and accurate measurements
  of intermittency},
  \href{https://doi.org/10.1103/physrevlett.123.014503}{\emph{Physical Review
  Letters} {\bfseries 123} (2019) }.

\bibitem{PhysRevLett.72.336}
Z.-S.~She and E.~Leveque, \emph{Universal scaling laws in fully developed
  turbulence}, \href{https://doi.org/10.1103/PhysRevLett.72.336}{\emph{Phys.
  Rev. Lett.} {\bfseries 72} (1994) 336}.

\bibitem{PhysRevE.63.026307}
V.~Yakhot, \emph{Mean-field approximation and a small parameter in turbulence
  theory}, \href{https://doi.org/10.1103/PhysRevE.63.026307}{\emph{Phys. Rev.
  E} {\bfseries 63} (2001) 026307}.

\bibitem{Eling2015TheAS}
C.~Eling and Y.~Oz, \emph{The anomalous scaling exponents of turbulence in
  general dimension from random geometry}, {\emph{Journal of High Energy
  Physics} {\bfseries 2015} (2015) 1}.

\bibitem{Oz:2017ihc}
Y.~Oz, \emph{{Spontaneous Symmetry Breaking, Conformal Anomaly and
  Incompressible Fluid Turbulence}},
  \href{https://doi.org/10.1007/JHEP11(2017)040}{\emph{JHEP} {\bfseries 11}
  (2017) 040} [\href{https://arxiv.org/abs/1707.07855}{{\ttfamily
  1707.07855}}].

\bibitem{Drygala_2022}
C.~Drygala, B.~Winhart, F.~di~Mare and H.~Gottschalk, \emph{Generative modeling
  of turbulence}, \href{https://doi.org/10.1063/5.0082562}{\emph{Physics of
  Fluids} {\bfseries 34} (2022) 035114}.

\bibitem{tretiak2022physicsconstrained}
D.~Tretiak, A.T.~Mohan and D.~Livescu, \emph{Physics-constrained generative
  adversarial networks for 3d turbulence},  2022.

\bibitem{Li2023}
T.~Li, M.~Buzzicotti, L.~Biferale and F.~Bonaccorso, \emph{Generative
  adversarial networks to infer velocity components in rotating turbulent
  flows}, \href{https://doi.org/10.1140/epje/s10189-023-00286-7}{\emph{The
  European Physical Journal E} {\bfseries 46} (2023) 31}.

\bibitem{Shu_2023}
D.~Shu, Z.~Li and A.B.~Farimani, \emph{A physics-informed diffusion model for
  high-fidelity flow field reconstruction},
  \href{https://doi.org/10.1016/j.jcp.2023.111972}{\emph{Journal of
  Computational Physics} {\bfseries 478} (2023) 111972}.

\bibitem{yang2023denoising}
G.~Yang and S.~Sommer, \emph{A denoising diffusion model for fluid field
  prediction},  2023.

\bibitem{fukami_fukagata_taira_2019}
K.~Fukami, K.~Fukagata and K.~Taira, \emph{Super-resolution reconstruction of
  turbulent flows with machine learning},
  \href{https://doi.org/10.1017/jfm.2019.238}{\emph{Journal of Fluid Mechanics}
  {\bfseries 870} (2019) 106–120}.

\bibitem{ZHOU2022105382}
Z.~Zhou, B.~Li, X.~Yang and Z.~Yang, \emph{A robust super-resolution
  reconstruction model of turbulent flow data based on deep learning},
  \href{https://doi.org/https://doi.org/10.1016/j.compfluid.2022.105382}{\emph{Computers
  \& Fluids} {\bfseries 239} (2022) 105382}.

\bibitem{li2023synthetic}
T.~Li, L.~Biferale, F.~Bonaccorso, M.A.~Scarpolini and M.~Buzzicotti,
  \emph{Synthetic lagrangian turbulence by generative diffusion models},  2023.

\bibitem{mohan2019compressed}
A.~Mohan, D.~Daniel, M.~Chertkov and D.~Livescu, \emph{Compressed convolutional
  lstm: An efficient deep learning framework to model high fidelity 3d
  turbulence},  2019.

\bibitem{king2018deep}
R.~King, O.~Hennigh, A.~Mohan and M.~Chertkov, \emph{From deep to
  physics-informed learning of turbulence: Diagnostics},  2018.

\bibitem{doi:10.1080/14685248.2020.1757685}
S.~Pandey, J.~Schumacher and K.R.~Sreenivasan, \emph{A perspective on machine
  learning in turbulent flows},
  \href{https://doi.org/10.1080/14685248.2020.1757685}{\emph{Journal of
  Turbulence} {\bfseries 21} (2020) 567}
  [\href{https://arxiv.org/abs/https://doi.org/10.1080/14685248.2020.1757685}{{\ttfamily
  https://doi.org/10.1080/14685248.2020.1757685}}].

\bibitem{moghaddam2018deep}
A.A.~Moghaddam and A.~Sadaghiyani, \emph{A deep learning framework for
  turbulence modeling using data assimilation and feature extraction},  2018.

\bibitem{li_yang_zhang_he_deng_shen_2020}
B.~Li, Z.~Yang, X.~Zhang, G.~He, B.-Q.~Deng and L.~Shen, \emph{Using machine
  learning to detect the turbulent region in flow past a circular cylinder},
  \href{https://doi.org/10.1017/jfm.2020.725}{\emph{Journal of Fluid Mechanics}
  {\bfseries 905} (2020) A10}.

\bibitem{Buzzicotti2022}
M.~Buzzicotti and F.~Bonaccorso, \emph{Inferring turbulent environments via
  machine learning},
  \href{https://doi.org/10.1140/epje/s10189-022-00258-3}{\emph{The European
  Physical Journal E} {\bfseries 45} (2022) 102}.

\bibitem{PhysRevFluids3104604}
P.~Clark Di~Leoni, A.~Mazzino and L.~Biferale, \emph{Inferring flow parameters
  and turbulent configuration with physics-informed data assimilation and
  spectral nudging},
  \href{https://doi.org/10.1103/PhysRevFluids.3.104604}{\emph{Phys. Rev.
  Fluids} {\bfseries 3} (2018) 104604}.

\bibitem{lellep_prexl_eckhardt_linkmann_2022}
M.~Lellep, J.~Prexl, B.~Eckhardt and M.~Linkmann, \emph{Interpreted machine
  learning in fluid dynamics: explaining relaminarisation events in
  wall-bounded shear flows},
  \href{https://doi.org/10.1017/jfm.2022.307}{\emph{Journal of Fluid Mechanics}
  {\bfseries 942} (2022) A2}.

\bibitem{whittaker2022neural}
T.~Whittaker, R.A.~Janik and Y.~Oz, \emph{Neural network complexity of chaos
  and turbulence},
  \href{https://doi.org/10.1140/epje/s10189-023-00321-7}{\emph{The European
  Physical Journal E} {\bfseries 46} (2023) 57}.

\bibitem{kohl2023turbulent}
G.~Kohl, L.-W.~Chen and N.~Thuerey, \emph{Turbulent flow simulation using
  autoregressive conditional diffusion models},  2023.

\bibitem{apte2023diffusion}
R.~Apte, S.~Nidhan, R.~Ranade and J.~Pathak, \emph{Diffusion model based data
  generation for partial differential equations},  2023.

\bibitem{lienen2023zero}
M.~Lienen, D.~Lüdke, J.~Hansen-Palmus and S.~Günnemann, \emph{From zero to
  turbulence: Generative modeling for 3d flow simulation},  2023.

\bibitem{Kolmogorov}
A.N.~Kolmogorov, \emph{The local structure of turbulence in incompressible
  viscous fluid for very large reynolds numbers}, {\emph{Cr Acad. Sci. URSS}
  {\bfseries 30} (1941) 301}.

\bibitem{ho2020denoising}
J.~Ho, A.~Jain and P.~Abbeel, \emph{Denoising diffusion probabilistic models},
  {\emph{Advances in neural information processing systems} {\bfseries 33}
  (2020) 6840}.

\bibitem{yang2023diffusion}
L.~Yang, Z.~Zhang, Y.~Song, S.~Hong, R.~Xu, Y.~Zhao et~al., \emph{Diffusion
  models: A comprehensive survey of methods and applications},
  \href{https://doi.org/10.1145/3626235}{\emph{ACM Comput. Surv.} (2023) }.

\bibitem{von-platen-etal-2022-diffusers}
P.~von Platen, S.~Patil, A.~Lozhkov, P.~Cuenca, N.~Lambert, K.~Rasul et~al.,
  ``Diffusers: State-of-the-art diffusion models.''
  \url{https://github.com/huggingface/diffusers}, 2022.

\bibitem{2dTurbReview}
G.~Boffetta and R.E.~Ecke, \emph{Two-dimensional turbulence},
  \href{https://doi.org/10.1146/annurev-fluid-120710-101240}{\emph{Annual
  Review of Fluid Mechanics} {\bfseries 44} (2012) 427}
  [\href{https://arxiv.org/abs/https://doi.org/10.1146/annurev-fluid-120710-101240}{{\ttfamily
  https://doi.org/10.1146/annurev-fluid-120710-101240}}].

\bibitem{specmeth}
C.~Canuto, M.~Hussaini, A.~Quarteroni and T.~Zang, \emph{Spectral methods.
  evolution to complex geometries and applications to fluid dynamics}, .

\bibitem{doi:10.1063/1.1762301}
R.H.~Kraichnan, \emph{Inertial ranges in two‐dimensional turbulence},
  \href{https://doi.org/10.1063/1.1762301}{\emph{The Physics of Fluids}
  {\bfseries 10} (1967) 1417}
  [\href{https://arxiv.org/abs/https://aip.scitation.org/doi/pdf/10.1063/1.1762301}{{\ttfamily
  https://aip.scitation.org/doi/pdf/10.1063/1.1762301}}].

\bibitem{PhysRevLett.81.2244}
M.A.~Rutgers, \emph{Forced 2d turbulence: Experimental evidence of simultaneous
  inverse energy and forward enstrophy cascades},
  \href{https://doi.org/10.1103/PhysRevLett.81.2244}{\emph{Phys. Rev. Lett.}
  {\bfseries 81} (1998) 2244}.

\bibitem{PhysRevE.82.016307}
G.~Boffetta and S.~Musacchio, \emph{Evidence for the double cascade scenario in
  two-dimensional turbulence},
  \href{https://doi.org/10.1103/PhysRevE.82.016307}{\emph{Phys. Rev. E}
  {\bfseries 82} (2010) 016307}.

\bibitem{loshchilov2019decoupled}
I.~Loshchilov and F.~Hutter, \emph{Decoupled weight decay regularization},  in
  \emph{International Conference on Learning Representations}, 2019,
  \href{https://openreview.net/forum?id=Bkg6RiCqY7}{https://openreview.net/forum?id=Bkg6RiCqY7}.

\bibitem{heyder2023generative}
F.~Heyder, J.P.~Mellado and J.~Schumacher, \emph{Generative convective
  parametrization of dry atmospheric boundary layer},  2023.

\end{thebibliography}\endgroup

\appendix
\section{Vorticity Notation}
\label{App:vort-vel}
In a two-dimensional, incompressible flow, the velocity components \( v_x \) and \( v_y \) can be described using a streamfunction \( \psi \) as:
\begin{align}
v_x = \frac{\partial \psi}{\partial y},  \quad & v_y = -\frac{\partial \psi}{\partial x} \ ,
\end{align}
and the vorticity, denoted by \( \omega \), is given by:
\begin{equation}
\omega = \frac{\partial v_y}{\partial x} - \frac{\partial v_x}{\partial y} = -\nabla^2 \psi.
\end{equation}
The kinetic energy per unit mass for an incompressible flow is given by:
\begin{equation}
E = \frac{1}{2} (v_x^2 + v_y^2) \ .
\end{equation}
Using the expressions for \( v_x \) and \( v_y \) from above:
\begin{align*}
E &= \frac{1}{2} \left( \left(\frac{\partial \psi}{\partial y}\right)^2 + \left(-\frac{\partial \psi}{\partial x}\right)^2 \right) \\
&= \frac{1}{2} \left( \left(\frac{\partial \psi}{\partial y}\right)^2 + \left(\frac{\partial \psi}{\partial x}\right)^2 \right).
\end{align*}

Using the definition of the Laplacian operator, \(\nabla^2 \psi = \frac{\partial^2 \psi}{\partial x^2} + \frac{\partial^2 \psi}{\partial y^2}\), and the expression for vorticity, we can express the energy spectrum \( E(k) \) in the wave number space as:
\begin{equation}
E(k) = \frac{1}{2k^2} |\omega(k)|^2.
\end{equation}

\end{document}